# Fullerene/antiferromagnetic reconstructed spinterface: subsurface layer dominates multi-orbitals' spin-splitting and large magnetic moment in $C_{60}$

Yangfan Shao[1,2], Rui Pang[3], Hui Pan[2,*], and Xingqiang Shi [1,*]

[1]*Department of Physics, South University of Science and Technology of China, Shenzhen 518055, China*

[2]*Institute of Applied Physics and Materials Engineering, Faculty of Science and Technology, University of Macau, Macau SAR, China*

[3]*School of Physics and Engineering, Zhengzhou University, Henan 450001, China*

\*E-mails: shixq@sustc.edu.cn (X.S.); huipan@umac.mo (H.P.)

**ABSTRACT**: The interfaces between organic molecules and metal surfaces with layered antiferromagnetic (AFM) order have gained increasing interests in the field of antiferromagnetic spintronics. The $C_{60}$/layered-AFM spinterfaces have been studied for $C_{60}$ bonded only to the outermost ferromagnetic layer. Using density functional theory calculations, here we demonstrate that $C_{60}$ adsorption can reconstruct the layered-AFM Cr(001) surface so that $C_{60}$ bonds to the top two Cr layers with opposite spin direction. Surface reconstruction drastically changes $C_{60}$'s spintronic properties: 1) the spin-split *p-d* hybridization involve multi-orbitals of $C_{60}$ and metal double layers, 2) the subsurface layer dominates the $C_{60}$ spin properties, and 3) reconstruction induces a large magnetic moment in $C_{60}$ of 0.58 $\mu_B$, which is a synergetic effect of the top two layers as a result of a magnetic direct-exchange interaction. Understanding these complex spinterfaces' phenomena is a crucial step for their device applications. The surface reconstruction can be realized by annealing at above room temperature in experiments.

Table of Contents Image:

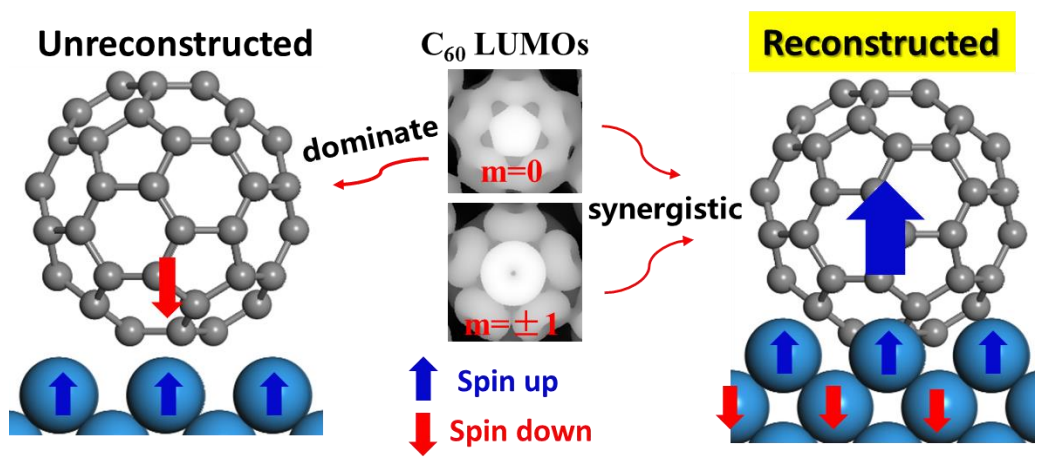

# INTRODUCTION

Antiferromagnetic (AFM) spintronics makes AFM properties useful and more interesting compared to traditional ferromagnetic (FM) spintronics devices[1-3]. This field of research has considerably grown and raised the hope for more efficient devices of data storage and computing[4-8]. Recently, the adsorption and coupling of molecules on surfaces with layered AFM order (interlayer in AFM order while intralayer in FM order) have been used to explore spin-polarized transport and the exchange bias effect[8-10]. A previous study proposed that a nanoscopic giant magnetoresistance (GMR) junction with a single molecule as the nonmagnetic spacer offers the possibility to replace one of the ferromagnetic electrodes by an antiferromagnet[10]. The significant advantage of the above structure is that the AFM materials are ideally hard magnetic[11-13], thus it is no longer necessary to use complex artificial antiferromagnets to pin the hard magnetic layer. Many works need to be done in order to understand and improve AFM spintronics. And it is necessary to develop novel spintronic device-models and practical antiferromagnetic materials.

Fullerenes $C_{60}$ are particularly promising for spintronic application due to the weak intrinsic spin-orbit and hyperfine interaction, resulting in long spin coherence time[14]. Several works have reported large magnetoresistance for $C_{60}$-based vertical spin valves[15-16]. Previous studies have shown that $C_{60}$ can induce the reconstruction of nonmagnetic metal surface such as Ag, Cu and Al[17-23], as well as the reconstruction of FM metal surfaces as Fe and Ni[24-26]. However, for $C_{60}$/AFM-metal interface systems, only unreconstructed (Unrec) adsorption structures have been reported[27-28]. It is unaware whether $C_{60}$ molecule can cause reconstruction on AFM metal surfaces. And if reconstructed, will the spintronic properties of adsorbed molecule change enormously?

Here we show that $C_{60}$ adsorption can induce the reconstruction of a layered AFM metal surface, Cr(001), using density functional theory calculations. We demonstrate that $C_{60}$ adsorption can induce reconstruction of Cr(001). The *p-d* hybridization between $C_{60}$ and Cr induces multi-orbitals' spin-split of $C_{60}$. With reconstruction (Rec), the m=$\pm$1 orbitals of $C_{60}$ LUMO[28] move closer to Fermi level and an inversion of $C_{60}$ spin-polarization occurs compared to the Unrec case. With Rec, the subsurface layer dominates the $C_{60}$ spin properties and the total magnetic moment induced in $C_{60}$ is enhanced enormously.

# COMPUTATIONAL METHODS

Most of the calculations adopt the plane-wave-basis-set Vienna ab initio simulation package

(VASP)[29-30]. Projector augmented wave potentials are employed with a kinetic energy cutoff of 500 eV[31]. The Perdew-Burke-Ernzerhof generalized gradient approximation is used for the exchange and correlation functional[27, 32]. A seven-layer slab of bcc-Cr(001) is used. The surface unitcell of $C_{60}$/Cr(001) adopts a quasi-hexagonal structure described by the following matrix notation[33]:

$$\begin{pmatrix} \vec{b_1} \\ \vec{b_2} \end{pmatrix} = \begin{pmatrix} 4 & 0 \\ 2 & 3 \end{pmatrix} \begin{pmatrix} \vec{a_1} \\ \vec{a_2} \end{pmatrix},$$

where $\vec{a_1}$, $\vec{a_2}$ and $\vec{b_1}$, $\vec{b_2}$ are the basis vectors of clean Cr(001) and $C_{60}$/Cr(001) lattices, respectively (see Figure 1a). The calculated Cr(100) surface lattice constant is 2.86 Å. The $C_{60}$/Cr(001) surface lattice can be described by a $4 \times \sqrt{13}$ supercell with the angle between two basis vectors of 56.3°. A 4×4 K-point sampling is utilized for the $4 \times \sqrt{13}$ surface supercell. The sign of the magnetic moment of the outermost Cr layer is defined positive and the sublayer negative. Calculations for spin-polarized density of states projected on $C_{60}$ molecular orbitals (MO-PDOS) utilized the Quantum Espresso package[34]. Energy cutoff of 50 and 500 Ry are employed for the wave functions and for charge density.

**RESULTS AND DISCUSSION**

**Adsorption Configuration.** Firstly, we search the adsorption structure with and without reconstruction. In order to find the most stable unreconstructed adsorption structure of $C_{60}$ on Cr(001) as shown in Figure 1a, $C_{60}$ adsorption with its hexagonal- and pentagonal-ring, and with a 6:6 double bond are considered; and, to find the optimum adsorption-site on Cr(001), we translate $C_{60}$ on the surface and rotate it around the surface normal. Hence, all the possible adsorption structures based on symmetry analysis of the two subsystems, $C_{60}$ and Cr(001), are considered [see Supporting Information]. Figure 1a shows the most stable unreconstructed configuration, $C_{60}$ bounds to Cr (001) by a pentagonal ring, which agrees well with literature[27]. A Cr atom is just below the pentagonal ring center with the shortest C-Cr bond length of 2.06 Å. This Cr atom bonds to $C_{60}$ the strongest and changes a lot in its magnetic moment, as shown in red in Figure 1c.

To find the most stable reconstructed configuration, removing one, four or five Cr atoms from the surface based on symmetry analysis are considered (Supporting Information). The size of a four atom hole in the top layer is suitable for one $C_{60}$ molecule to sink in, and larger hole is not necessary. Figure 1b shows the most stable reconstructed configuration with a 4-atom hole. Two interlayer magnetic-orderings between the outermost- and sub-layers of Cr, AFM and FM ordering are considered for the Rec case. The AFM ordering is lower in energy by 1.71 eV per supercell. So the top two layers remain in AFM ordering with Rec.

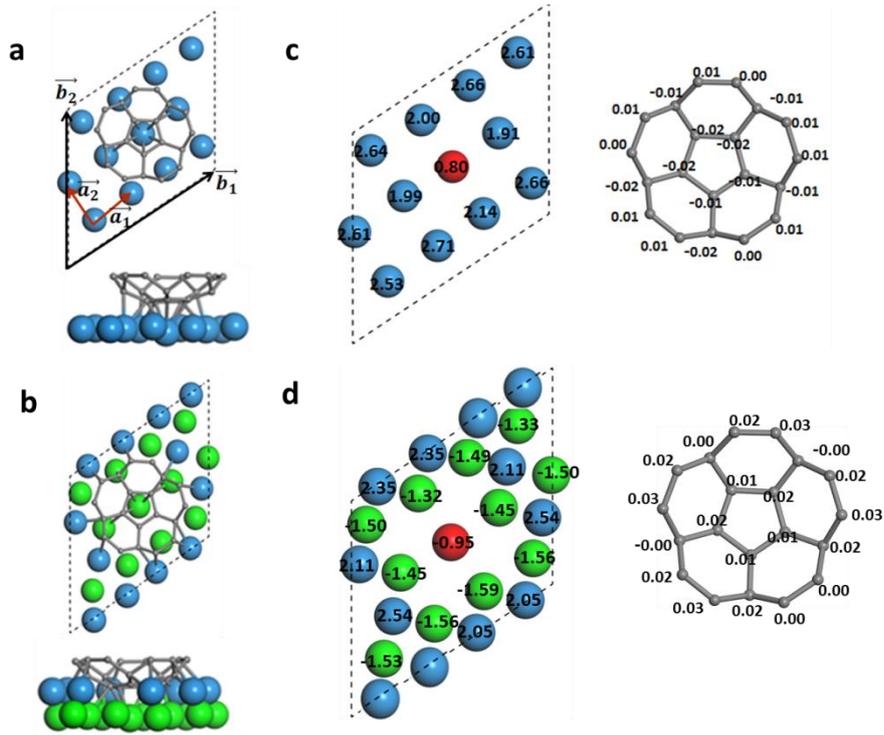

Figure 1. Top and side views of the unreconstructed (Unrec) structure with $C_{60}$ pentagon on Cr(001) (a); and the reconstructed (Rec) structure with $C_{60}$ pentagon sinks-in a 4-atom hole of Cr(001) (b). Magnetic moments (in $\mu_B$) from Bader charge analysis at the $C_{60}$/Cr(001) interfaces for Cr(001) surface layers and for $C_{60}$ bottom part in the Unrec (c) and Rec (d) structures. The "red" atom in (c) and (d) denotes the Cr atom just below $C_{60}$ pentagon in (a) and (b), respectively.

To determine the relative energetic stability between Unrec and Rec structures, we compare the energy gain from forming more C-metal bonds in the Rec structure and the energy cost from forming a several-atom hole (the surface vacancy formation energy $E_{vac}$). The kinked-edge model[35] is adopted to calculate $E_{vac}$, with the missing atoms below $C_{60}$ moving to the surface kinked-edge site (see Supporting Information). The adsorption energy corrected for $E_{vac}$ is expressed as $E_{ads} = E_{C60/Cr(001)} - E_{C60} - E_{Cr(001)} + E_{vac}$, where $E_{C60/Cr(001)}$, $E_{C60}$ and $E_{Cr(001)}$ are energies of the $C_{60}$/Cr(001) full system, the isolated $C_{60}$ and Cr(001) surface, respectively. It is −4.19 eV for the Unrec structure and is −4.84 eV for the Rec with a 4-atom hole, which demonstrates that the Rec structure is much more energetically stable. Note that previous experiments reported that surface reconstructions already occur for annealing temperatures in the range of 340-570 K for various

surfaces with $C_{60}$[23, 33, 36-39].

**Electronic and Magnetic Properties.** We turn to discuss electronic and magnetic properties of C60 molecule with and without reconstructions. Based on the above analysis, the most stable reconstructed structure is a $C_{60}$ pentagon sinks in a surface 4-atom hole. Nineteen C-Cr bonds are formed in total (for C-Cr bonds less than 2.20 Å); and, in the first layer the eight surrounding Cr atoms are all bonded with C. Thus, in contrast to the unreconstructed structure, the interaction between $C_{60}$ and Cr (001) becomes much stronger with Rec. Surface reconstruction drastically changes $C_{60}$'s spintronic properties, as detailed below.

Figure 2 compares the spin-polarized properties of $C_{60}$ with and without reconstruction. Figure 2a shows that, compared to isolated $C_{60}$, the adsorbed $C_{60}$ orbitals significantly broadened and shifts-down in energy because of hybridization with Cr *d*-electrons and spin-polarized charge transfer from the Cr surface to $C_{60}$. One can see spin-dependent energy level shifting and broadening from Figure 2a. To show these characters clearly, Figures 2c and 2d shows the DOS projected on $C_{60}$ highest-occupied molecular orbitals (HOMO) and lowest-unoccupied molecular orbitals (LUMO), for the Unrec and Rec cases, respectively. The broadened HOMO and LUMO (m=0 and $\pm 1$) in the Rec case extends to even lower energies than that in Unrec. Around Fermi level, the DOS projected onto $C_{60}$ LUMOs dominate the spin properties of the adsorbed molecule. For adsorbed $C_{60}$, the threefold degenerate LUMO levels split into m=0 and $\pm 1$. The three LUMO states have different spatial distributions. The m=0 orbital is localized at the pentagon center and the m=$\pm 1$ orbitals are localized on the pentagon ring edges. The simulated *STM* maps at energy range of -0.4 ~ 0 eV for spin down and 0.8~1.2 eV, showing the signatures of LUMO m=0, and m=±1 molecular orbitals in Figure 2c. Due to the adsorption-induced symmetry breaking, the m=0 and m=$\pm 1$ orbitals split in different manners for the Unrec and Rec cases. The m=$\pm 1$ states in the Rec is much closer to Fermi level compared to the Unrec case. The HOMO is further down-shifted from Fermi level by about 0.5 eV in Rec in contrast to the Unrec case. Part of HOMO states transfer to above the Fermi level, especially with Rec.

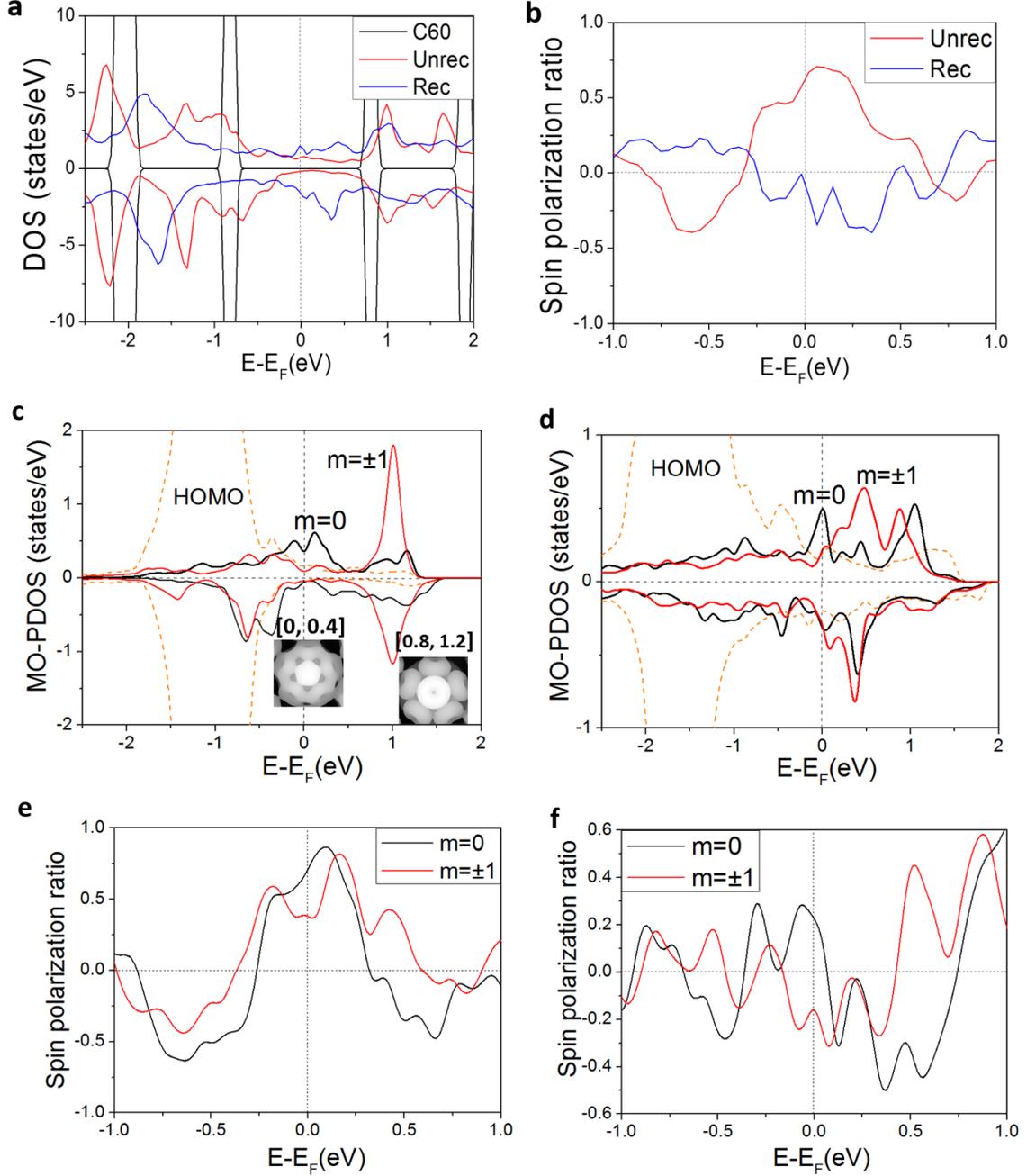

Figure 2. (a) Spin-polarized density of states (DOS) of $C_{60}$ in the Unrec (Figure 1a) and Rec (Figure 1b) structures and of gas phase $C_{60}$. (b) Spin-polarization ratio (SPR) of $C_{60}$ in the Unrec and Rec cases. DOS projected on $C_{60}$ molecular orbitals (MO-PDOS) for HOMO and LUMO (m=0, ±1) for the Unrec (c) and Rec (d) cases. SPR for m=0, ±1 for the Unrec (e) and Rec (f) cases. Inset in (c): simulated $dI/dV$ maps at the energy range of [0, 0.4] eV of spin down and [0.8, 1.2] eV of both spin, showing the signatures of m=0 and m=±1 orbitals, respectively.

Figure 2b shows that the $C_{60}$ spin-polarization with Rec is inversed (opposite sign) compared to the Unrec case. In order to quantify the effect on spin polarization of $C_{60}$ molecule, we calculate the spin polarization ratio (SPR). The spin-polarization ratio (SPR) is defined as SPR(E) = [DOS$_\uparrow$(E) − DOS$_\downarrow$(E)] / [DOS$_\uparrow$(E) + DOS$_\downarrow$(E)], where DOS$_\uparrow$(E) and DOS$_\downarrow$(E) are the PDOS for the majority-spin and minority-spin, respectively. For Unrec, the m=0 state of LUMO plays a dominant contribution to spin-polarization around Fermi level (Figures 2c and 2e). However, for the Unrec case, although the m=0 of LUMO state is still important, the m=±1 states dominate the spin-polarization around Fermi-level (Figures 2d, 2f and 2b). Note that the sum of the m=±1 states is divided by a factor of two. The m=±1 states move-down nearer to Fermi level with Rec. The spin-polarization inversion occurs for the m=±1 states in Rec relative to Unrec (Figure 2f) since the subsurface layer with an opposite spin dominates the C$_{60}$ spin-polarization. The total spin-polarization inversion with Rec as shown in Figure 2b is a hint for sublayer domination. To prove it (sublayer domination), we take only the top Cr layer and C$_{60}$, keeping it in the Rec structure and calculate the C$_{60}$ SPR -- a positive smaller SPR than the Unrec structure is obtained (Supporting Information). One should mention that density functional theory (DFT) calculations overestimate the magnetic moment of the outermost Cr layer[27]. Even though, there is still an inversion of SPR with Rec due to the sublayer domination. This means that stronger p-d coupling between C$_{60}$ and subsurface metal atoms (especially the "red" Cr atom just below C$_{60}$ pentagon ring shown in Figure 1d) causes the inversion of C$_{60}$ spin polarization.

Figure 2f shows that the SPR of m=±1 states is inversed in Rec relative to Unrec in Figure 2e, because the subsurface-layer with an opposite-spin dominates the C$_{60}$ spin-polarization. However, for the m=0 states the SPR is still positive around Fermi level, which is explained below in Figure 3.

Figures 3a and 3c present out-of-plane d–orbitals of the "red" Cr atom since there is a strong coupling between C$_{60}$ p$_z$ and out-of-plane d–orbitals of Cr. Figures 3a, 3b and 2c shows that for the Unrec case there's a spin-dependent energy shifting of the m=0 state, which has been discussed in detail in references[27, 40]. Figures 3c, 3d and 2d shows that with Rec, when the C$_{60}$ energy-level broadening is far greater than the energy different between C$_{60}$ m=0 level and Fermi level, an inversion of the spin polarization on the m=0 molecular orbital is induced at Fermi level. The mechanism is detailed in literature[40] by assuming that the spin-dependent molecular-level broadening is in direct proportion to the amount of Cr spin-dependent DOS at Fermi level -- for the peak of m=0 at Fermi level, the broadening of spin-up states is smaller and hence at Femi level the spin-polarization of m=0 is inversed relative to that of the sublayer Cr atom.

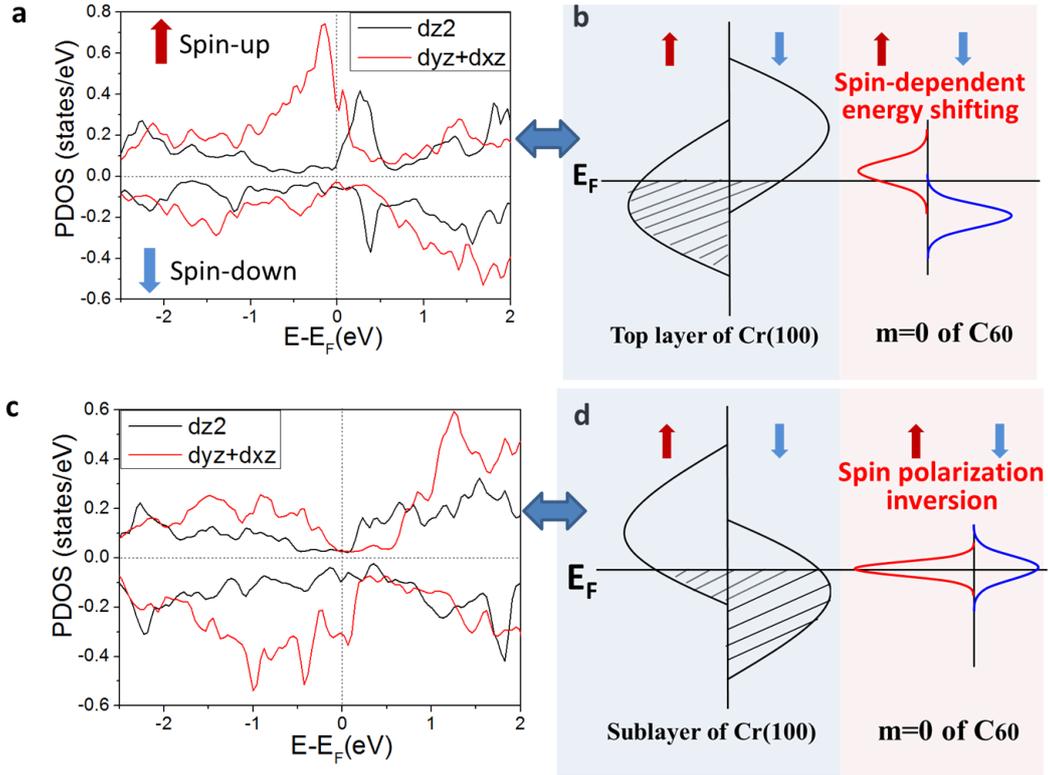

Figure 3. Outer-of-plane *d*-orbital PDOS of the "red" Cr atom in Figures 1c and 1d in the Unrec (a) and Rec (c) cases. (b) Sketch map indicates the coupling between Cr states in (a) and $C_{60}$ m=0 states in Figure 2c for the Unrec case; and (d) sketch map indicates the coupling between Cr states in (c) and $C_{60}$ m=0 states in Figure 2d for the Rec case.

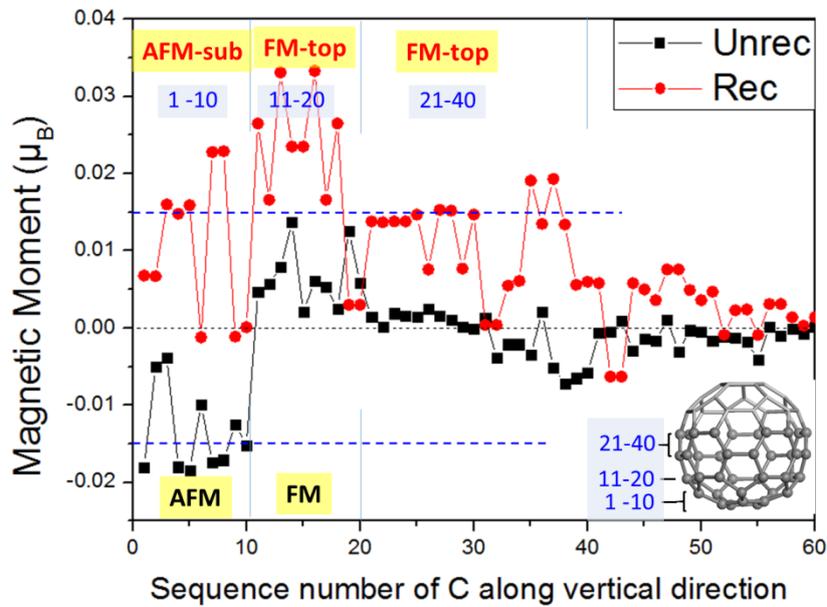

Figure 4. The magnetic moments (in $\mu_B$) of C atoms along the vertical-direction in the Unrec and Rec cases.

We calculate the magnetic moment using Bader charge analysis[41-42]. In Unrec, the magnetic moment of the "red" Cr atom just below the pentagon ring is 0.80 $\mu_B$ (Figure 1c), which is much smaller than that of other Cr atoms. The magnetic moments are also reduced for the four Cr atoms surrounding the "red" Cr atom. In Rec, the magnetic moment of the subsurface "red" Cr below $C_{60}$ pentagon is reduced to $-0.95$ $\mu_B$ (Figure 1d). The total magnetic moment in $C_{60}$ is 0.58 $\mu_B$ for Rec while it is -0.11$\mu_B$ for Unrec. So, in both Rec and Unrec cases, the induced magnetic moments in $C_{60}$ are in an AFM-like coupling to the surface layer that contains the "red" Cr atom -- namely, the outermost layer in Unrec and the sublayer in Rec. This spin character induced in $C_{60}$ is shown more clearly in Figure 4. For simplicity, one can focus only on the C atoms with the magnitude of magnetic moments close to or larger than 0.015 $\mu_B$. Under this simplification, the induced magnetic moment in $C_{60}$ can be explained by the magnetic direct-exchange mechanism, which results in AFM-like coupling between Cr-C for smaller Cr-C distances while FM-like coupling for larger Cr-C distances. In the Unrec case, the nearest Cr-C distance is 2.06 Å, Figure 4 shows that there's a AFM-like coupling between the Cr surface and the nearest ten C atoms (the C atoms labeled by 1-10 in Figure 4), while the coupling change to FM-like for the second-nearest-neighbor ten C atoms (labeled by 11-20). In the Rec structure, AFM-like coupling between Cr-C also appears for the ten C atoms close to the sublayer (labeled by AFM-sub in Figure 4). However, FM-like coupling between Cr-C appears for the other C atoms with the top-layer (labeled by FM-top in Figure 4), due to the nearest Cr-C distance for the top-layer are larger than that for the sublayer (2.15 Å for top-layer compared to 2.06 Å for sublayer). Figure 4 shows that there is an enhancement of magnetic moment in the Rec case. The number of transferred electrons to $C_{60}$ is also enhanced for Rec. The number of transferred electrons is 2.83 *e* for Unrec and 4.00 *e* for Rec, respectively.

## SUMMARY

In summary, we have theoretically investigated $C_{60}$ on layered AFM Cr(001), both with and without reconstruction. The *p-d* interaction between $C_{60}$ and Cr induces the multi-orbitals' spin-splitting of LUMO states around Fermi level. For the unreconstructed case, the LUMO m=0 dominates the spin properties around Fermi level; while for the reconstructed case, all three LUMOs play a significant role. More importantly, in the reconstructed case the subsurface layer dominates the $C_{60}$ spin properties due to stronger *p-d* coupling of $C_{60}$ with the sublayer. In the reconstructed case, a large magnetic moment is induced in $C_{60}$ of 0.58 $\mu_B$, which is a synergetic effect of the top two layers as a result of a magnetic direct-exchange interaction.

## ASSOCIATED CONTENT

**Supporting Information**

Additional information about structure determination and spintronic properties.

## ACKNOWLEDGMENTS

This work is supported by the National Natural Science Foundation of China (Grants 11474145 and 11334003), the Young Teachers Special Startup Funds of Zhengzhou University, the Science and Technology Development Fund from Macau SAR (FDCT-068/2014/A2, FDCT-132/2014/A3, and FDCT-110/2014/SB) and Multi-Year Research Grants (MYRG2014-00159-FST and MYRG2015-00017-FST) from Research & Development Office at University of Macau.